\documentclass[a4paper]{article}

\usepackage[english]{babel}
\usepackage[utf8x]{inputenc}
\usepackage{amsmath}
\usepackage{graphicx}
\usepackage[colorinlistoftodos]{todonotes}
\usepackage{amssymb}
\makeatletter \renewcommand\@biblabel[1] \makeatother

\title{Algorithmic Trading with Fitted Q Iteration and Heston Model}
\author{Son Le}

\begin{document}
\maketitle

\begin{abstract}
We present the use of the fitted Q iteration in algorithmic trading. We show that the fitted Q iteration helps alleviate the dimension problem that the basic Q-learning algorithm faces in application to trading. Furthermore, we introduce a procedure including model fitting and data simulation to enrich training data as the lack of data is often a problem in realistic application. We experiment our method on both simulated environment that permits arbitrage opportunity and real-world environment by using prices of 450 stocks. In the former environment, the method performs well, implying that our method works in theory. To perform well in the real-world environment, the agents trained might require more training (iteration) and more meaningful variables with predictive value.
\end{abstract}

\section{Introduction}

In this paper, we investigate the use of Fitted Q Iteration and Heston Model in trading. Ritter (2017) illustrates the use of Q-Learning in trading by using a simulated path of a Ornstein-Uhlenbeck process. The Q-learner in Ritter's study is able to find a policy that performs superbly for the Ornstein-Uhlenbeck process. However, to train a Q-learner using real-world data, one needs a big amount of data; and finding such an amount might not be feasible. \

We try to assuage the data issue by extending Ritter's idea by performing a multi-step training procedure: (1) assume a reasonable stochastic process model for asset returns, (2) estimate the parameters of the model from market data, (3) use the model to simulate a much larger data set than the real world presents, and (4) train the reinforcement-learning system on the simulated data. Section 2 discusses the Heston Model, which is the model of our choice, as well as parameter estimation of the Heston Model by using the extended Kalman filter and pseudo maximum likelihood estimation.

Furthermore, while the standard table Q-learning converges for a discrete \textit{and} small action space \textit{and} state space, the same convergence may not apply for data with continuous values, big action space, and big state space. Therefore, we are going to experiment an extension of Q-learning, a method known as fitted Q iteration in order to improve this shortcoming. An introduction of the fitted Q iteration algorithm will be in section 3.

In section 4, we are going to perform a numerical experiment using the same trading mechanism found in Ritter (2017) with market data. Out of sample result of the Fitted Q agent will be included in section 4.

\section{Heston Model and Parameter Estimation}
\subsection{Heston Model}
The Heston model in a risk-neutral probability space $\mathbb{Q}$ (Shreve, 2014) is presented as follow:
\begin{equation}
\begin{aligned}
dS(t) &= rS(t)dt + \sqrt{V(t)}S(t)dW_1(t), \\
dV(t) &= \kappa(\theta-V(t))dt + \sigma\sqrt{V(t)}dW_2(t), \\
dW_1&(t)dW_2(t) = \rho dt, \ \ t \in [0,T]
\end{aligned}
\end{equation}
where $S(t)$ is the price of a stock at time $t$ and $V(t)$ is the variance of the price. For convenience of notion, following other studies on volatility, we will regard $V(t)$ as volatility. $[0,T]$ is the time interval. And $r$ is the risk-free interest rate.

The equation for volatility $V(t)$ is a square root mean-reversion model, which implies the mean-reversion property of market volatility. $\theta$ therefore is interpreted as the long run mean of volatility and $\kappa$ is the mean reversion rate. $\sigma$ is the volatility of volatility and influences the variation of the volatility and kurtosis of the stock return rate. Cox et al. (1985) mentions that if $2\kappa \theta > \sigma^2$, the volatility is always above zero and thus ensure the positiveness of volatility. Lastly, $W_1(t)$ and $W_2(t)$ are Brownian motions with correlation coefficient $\rho$ in the risk neutral measure $\mathbb{Q}$. Coefficient $\rho$ affects the heavy tails of the stock return rate distribution.

There are other models that can be used to estimate a stock process such as the geometric Brownian motion (Black \& Scholes, 1973) and the Hull-White model (Hull \& White, 1987). In this paper, we choose the Heston model because it assumes that volatility is non-constant, in contrast to the Black Scholes model; and that it restricts volatility from being negative, in contrast to the Hull-White model. However, one can certainly experiment with other models to test if they performs better than the Heston model in specific cases.

One can get the stock price process $S(t)$ from the market data. However, the data on volatility $V(t)$ process is unavailable. Estimating parameters of the Heston model therefore is difficult since it is not possible to implement a straightforward Maximum Likelihood Estimation. Furthermore, there is no closed-form solution for the Heston model. There are several methods to estimate parameters of the Heston model such as the extended Kalman filter (Javaheri et al., 2003), the unscented Kalman filter (Li, 2013), and the consistent extended Kalman filter (Jiang et al., 2014). In this paper, we are using a method known as psuedo-maximum likelihood estimation and extended Kalman filter to estimate the parameters of the Heston model (Wang et al., 2017; Javaheri et al., 2003).

\subsection{Parameter Estimation}
\subsubsection{ The Extended Kalman Filter}
First, let's apply Ito's lemma on ln$S(t)$ to get the equation of rate of return on the stock $S_t$:
\begin{equation}
\begin{aligned}
d(\text{ln}S(t)) &= \frac{1}{S(t)}dS(t)-\frac{1}{2S^2(t)}dS(t)dS(t) \\
&= (r-\frac{1}{2}V(t))dt + \sqrt{V(t)}dW_1(t)
\end{aligned}
\end{equation}
We then can obtain another form of the Heston model:
\begin{equation}
\begin{aligned}
dV(t) &= \kappa(\theta-V(t))dt + \sigma\sqrt{V(t)}dW_v(t), \\
d(\text{ln}S(t)) &= (r-\frac{1}{2}V(t))dt + \sqrt{(1-\rho^2)V(t)}dW_s(t) +\rho\sqrt{V(t)}dW_v(t)
\end{aligned}
\end{equation}
where
$$dW_s(t)dW_v(t)=0$$
and 
\begin{equation}
\begin{aligned}
W_1(t) &= \sqrt{1-\rho^2}W_s(t) + \rho W_v(t)\\
W_2(t) &= W_v(t)
\end{aligned}
\end{equation}
To introduce the extended Kalman filter, we discretize the Heston model equation (2.3) to get:
\begin{equation}
\begin{aligned}
V_k &= V_{k-1} + \kappa(\theta-V_{k-1})\Delta t + \sigma \sqrt{V_{k-1}} \Delta W_{vk} \\
\text{log}(S_{k+1}) &= \text{log}{S_k} + (r-\frac{1}{2}V_k)\Delta t + \sqrt{(1-\rho^2)V_k} \Delta W_{sk} + \rho \sqrt{V_k}\Delta W_{vk}
\end{aligned}
\end{equation}
where $V_k$ is volatility of stock price at discretized step $k$, and $\Delta W_{sk}$ and $\Delta W_{vk}$ are Brownian increments.

Because the system (5) is nonlinear, we can use the extended Kalman filter, which linearizes a nonlinear system to a linear system, to estimate the unknown volatility (Javahari et al., 2003; Simon, 2006). We can write down the extended Kalman filter algorithm for the Heston model as follow:

\textbf{(1) Initialization:}
\begin{equation}
\begin{aligned}
\hat{V}_0 &= E(V_0) \\
P_0 &= E[(V_0-\hat{V}_0^{+})(V_0-\hat{V}_0^{+})^T] \\
\Gamma_0 &= (\hat{\kappa}_0,\hat{\theta}_0,\hat{\sigma}_0,\hat{\rho}_0)^T 
\end{aligned}
\end{equation}

For k in 1...N:

\textbf{(2) Linearization of the state equation:}
\begin{equation}
\begin{aligned}
F_k &= 1 - \hat{\kappa}_k \Delta t \\
L_k &= \Big[ 0, \hat{\sigma}_k\sqrt{\hat{V}_k \Delta t} \Big]
\end{aligned}
\end{equation}

\textbf{(3) Update state prediction estimation and prediction estimation error covariance:}
\begin{equation}
\begin{aligned}
\bar{V}_{k+1} &= \hat{V}_k + \hat{\kappa}_k\hat{\theta}_k\Delta t - \hat{\kappa}_k \hat{V}_k \Delta t \\
\bar{P}_{k+1} &= F_k P_k F_k^T + L_k Q_k L_k^T
\end{aligned}
\end{equation}

\textbf{(4) Linearization of the measurement function:}
\begin{equation}
\begin{aligned}
H_{k+1} &= -\frac{1}{2}\Delta t \\
M_{k+1} &= \Big[ \sqrt{(1-\hat{\rho}_k^2)\bar{V}_{k+1}\Delta t}, \hat{\rho}_k \sqrt{\bar{V}_{k+1}\Delta t} \Big]
\end{aligned}
\end{equation}

\textbf{(5) Update state estimate and estimation error covariance:}
\begin{equation}
\begin{aligned}
K_{k+1} &= (\bar{P}_{k+1}H_{k+1}^T + L_k Q_k M_{k+1}^T)(H_{k+1}\bar{P}_k H_{k+1}^T + \\&M_{k+1} Q_k M_{k+1}^T + H_{k+1} L_k Q_k M_{k+1}^T + M_{k+1} Q_k L_k^T H_{k+1}^T)^{-1} \\
\hat{V}_{k+1} &= \bar{V}_{k+1} +  K_{k+1}\Big[ \text{log}S_{k+1} - \text{log}S_{k} - \frac{1}{2}\bar{V}_{k+1} \Big] \\
P_{k+1} &= \bar{P}_{k+1} - K_{k+1}(H_{k-1}\bar{P}_{k+1} + M_{k+1}L_k^T)
\end{aligned}
\end{equation}

\textbf{(6) Conduct parameter estimation}
\subsubsection{Pseudo-Maximum Likelihood Estimation}
The usage of a pseudo-maximum likelihood estimation is outlined in Wang et al., (2017). The goal of the method is obtaining the set of parameters in the Heston model. For the state equation, an approximation of the volatility process (Nowman, 1997) can be written as follow:
\begin{equation}
dV(t) = \kappa(\theta - V(t - \Delta t))dt + \sigma\sqrt{V(t-\Delta t)} dW_2(t)
\end{equation}
where $V(t)$ remains constant in small interval $[t-\Delta t,t)$. The approximation of the above equation (11) is:
\begin{equation}
V(t) = e^{-\kappa \Delta t}V(t-\Delta t) + \theta(1-e^{-\kappa \Delta t}) + \xi_t
\end{equation}
where $E(\xi_t)=0$, $E(\xi_t \xi_s) = 0$ when $t \neq s$, and $E(\xi_t^2)=\frac{1}{2}\sigma^2 \kappa^{-1}(1-e^{-2\kappa \Delta t})V(t-\Delta t)$. Then, assuming that $\xi_t$ follows a Gaussian distribution, the variance of $\xi_t$ is:
\begin{equation}
\text{Var}(\xi_t;V(t-\Delta t), \kappa, \theta, \sigma) = \frac{1}{2}\sigma^2 \kappa^{-1}(1-e^{-2\kappa \Delta t})V(t-\Delta t)
\end{equation}
Now, we can discretize the process $V(t)$ as $\{V_0,V_{\Delta t},...,V_{n\Delta t}\}$. Denoting $n\Delta t$ as $k$, we can write down the pseudo log-likelihood function of $\xi_t$ as follow:
\begin{equation}
l_v(\kappa,\theta,\sigma) = - \sum_{k=1}^n \Big[ \frac{1}{2} \text{log}(\text{Var}(\xi_t;V(t-\Delta t)) + \frac{1}{2}\text{Var}^{-1}(\xi_t;V(t-\Delta t))( V_k - e^{-\kappa \Delta t V_{k-1} - \theta(1-e^{-\kappa \Delta t}) } ) \Big]  
\end{equation}
Taking derivative of the above likelihood function (14), we can obtain formula for the parameters (Tang \& Chen, 2009):
\begin{equation}
\begin{aligned}
\hat{\kappa} &= -\frac{1}{\Delta t}\text{log}(\hat{\beta}_1) \\
\hat{\theta}&=\hat{\beta}_2 \\
\hat{\sigma}^2 &= \frac{2\hat{\kappa}\hat{\beta}_3}{1-\hat{\beta}_1^2}
\end{aligned}
\end{equation}
where
\begin{equation}
\begin{aligned}
\hat{\beta}_1 &= \frac{n^{-2}\sum_{k=1}^{n}V_k\sum_{k=1}^nV_{k-1}^{-1}-n^{-1}\sum_{k=1}^n V_k V_{k-1}^{-1}}{n^{-2}\sum_{k=1}^n V_{k-1} \sum_{k=1}^n V_{k-1}^{-1} - 1} \\ 
\hat{\beta}_2 &= \frac{n^{-1}\sum_{k=1}^n V_k V_{k-1}^{-1} - \hat{\beta}_1}{(1-\hat{\beta}_1)n^{-1}\sum_{k=1}^nV_{k-1}^{-1}} \\
\hat{\beta}_3 &= n^{-1} \sum_{k=1}^n(V_k-V_{k-1}\hat{\beta}_1 - \hat{\beta}_2(1-\hat{\beta}_1)^2) V_{k-1}^{-1}
\end{aligned}
\end{equation}
Rewriting the asset return equation (measurement equation) in (3), we get:
\begin{equation}
d\text{ln}(S(t)) = (r-\frac{1}{2}V(t))dt + \sqrt{(1-\rho^2)V(t)}dW_s(t) + \frac{\rho}{\sigma}(dV(t)-\kappa(\theta-V(t))dt)
\end{equation}
Discretizing the above equation and substituting $V(t)$ into the equation, we obtain:
\begin{equation}
\begin{aligned}
\text{ln}(S(t)) = \text{ln}(S(t-\Delta t) + &\Big[ r-\frac{1}{2}V(t-\Delta t) - \frac{\rho}{\sigma} \kappa(\theta- V(t-\Delta t)) \Big] \Delta t   \\
+ &\frac{\rho}{\sigma} (V(t) - V(t-\Delta t)) + \eta_t
\end{aligned}
\end{equation}
where $E(\eta_t)=0$, $E(\eta_t\eta_s) = 0$ when $t \neq s$, and $E(\eta_t^2)=V(t-\Delta t)(1-\rho^2)\Delta t$. Then, similarly, assuming that $\eta_t$ follows a Gaussian distribution, the variance of $\eta_t$ is:
\begin{equation}
\text{Var}(\eta_t;V(t-\Delta t),\rho)=V(t-\Delta t)(1-\rho^2)\Delta t
\end{equation}
We can write the pseudo log-likelihood (Wang et al., 2017) from the discrete form of the measurement function as follow:
\begin{equation}
l_s(\rho) = - \sum_{k=1}^n \Big[ \frac{1}{2} \text{log}(\text{Var}(\eta_t;V_{k-1},\rho)) + \frac{1}{2}\text{Var}^{-1}(\eta_t;V_{k-1},\rho)( m^2(\eta_t;V_{k-1}) \Big] 
\end{equation}
where 
\begin{equation}
m(\eta_t;V_{k-1}) = \text{ln}(S_k) - \text{ln}(S_{k-1}) + \Big[ r - \frac{1}{2}V_{k-1} - \frac{\rho}{\sigma}\kappa (\theta-V_{k-1}) \Big] \Delta t + \frac{\rho}{\sigma}(V_k - V_{k-1})
\end{equation}
Maximizing the above log-likelihood function, we can obtain $\hat{\rho}$ i.e:
\begin{equation}
\hat{\rho} = \text{argmin} (-l_s(\rho))
\end{equation}
We then can use equations (15) and (22) for step 6 in the extended Kalman filter algorithm outlined in the last subsection and complete the parameter estimation process.

\section{Fitted Q Iteration}
Research in reinforcement learning aims at designing algorithms by which autonomous agents learn to behave from their interaction with the environment to gather information and yield the optimal policy. The work of Sutton \& Barto (1998) covers well the theoretical details of reinforcement learning. 

Mathematically, in a reinforcement learning problem, the goal of the agent is to maximize the expected cumulative reward:
\begin{equation}
G_T = R_{t+1} + \lambda R_{t+1} + \lambda^n R_{t+3}+ ...
\end{equation}
by searching for a policy that maximizes $E[{G_t}]$. The constant $\lambda \in [0,1]$ is the discount rate, and is useful in problems where $T=\infty$. The key idea of reinforcement learning is the use of value functions to organize and structure the search for good policies (Sutton \& Barto, 1998). The state-value function for a policy $\pi$ is denoted as:
\begin{equation}
v_\pi(s) = E_\pi[G_t|S_t=s]
\end{equation}
where $E_\pi$ is the expectation of reward if policy $\pi$ is followed. Similarly, the action-value function expresses the value of starting in state $s$, taking action $a$, and then following policy $\pi$ thereafter:
\begin{equation}
q_\pi(s,a) = E_\pi[G_t|S_t=s,A_t=a]
\end{equation}
A policy $\pi$ is defined to be at least at good at any other policy $\pi$' if $v_\pi(s)\geq v_{\pi '}(s)$ for all $s$; and an optimal policy is one that is at least as good as any other policy (Ritter, 2017). Then the optimal state-value function $v^*(s) = \text{max}_\pi v_\pi$ and optimal action-value function $q^*=\text{max}_\pi q_\pi (s,a)$ satisfy the Bellman optimality equation:
\begin{equation}
\begin{aligned}
v^* &= \text{max}_a \sum_{s',r} p(s',r|s,a)[r+\lambda v^*(s')] \\
q^*(s,a) &= \sum_{s',r} p(s',r|s,a)[r+\lambda \text{max}_{a'}q^*(s',a')]
\end{aligned}
\end{equation}
where the sum over $s',r$ denotes a sum over all states $s'$ and all rewards $r$. Then if we can get a function $q(s,a)$ that estimates $q^*(s,a)$ then the greedy is defined as picking at time $t$ an action $a_t^*$ that maximizes $q(s_t,a)$ over all possible a, where $s_t$ is the state at time $t$.

Watkins (1989) made a breakthrough in reinforcement learning as the author suggested an iterative method that converges to the optimal action-value function $q^*$. The algorithm is as follow:

\textbf{(1) Initialization:}
Initialize a matrix $Q$ with one row per state, and one column per action. It can be the zero matrix or available prior information.

Until converged or some criteria is met, do:

\textbf{(2) Choose an action} $A$ \textbf{using a policy derived from} $Q$

\textbf{(3) Take the action} $A$ \textbf{and go to a new state} $S'$ \textbf{where a reward} $R$ \textbf{is observed}

\textbf{(4) Update the value of} $Q(S,A$):
\begin{equation}
Q(S,A) \leftarrow Q(S,A) + \alpha[R+\lambda \text{max}_{a} Q(S',a) - Q(S,A)]
\end{equation}
where $\alpha \in (0,1)$ is a step-size parameter that influences the rate of learning.

However, in dealing with continuous or very large state and/or action spaces, the $Q$-function cannot be represented as a matrix like above because it would either not converge or run out of storage space. To overcome this problem, Ormoneit \& Sen (2002) introduced kernel-based reinforcement learning by reformulating the $Q$-function determination problem as a sequence of kernel-based regression problems. This framework allows for the generalization capability to use any regression algorithm to estimate the $Q$-function. The term \textit{fitted Q iteration} was coined by Ernst et al., (2003).

Different from the traditional table $Q$-learning, the fitted $Q$ iteration is a batch mode reinforcement learning algorithm. The fitted $Q$ iteration yields an approximation of the $Q$-function corresponding to an infinite horizon optimal control problem with discounted rewards, by iteratively extending the optimization horizon (Ernst et al., 2003).

Theories and proof of convergence of the fitted Q iteration algorithm can be found in (Ernst et al., 2003). The outline of the algorithm is as follow:

\textbf{(1) Prepare training set:} By running simulations, prepare the training set of four-tuple: \{$S_t , A_t, R_t , S_{t+1}$\} where $S_t$ is the state at time $t$, $A_t$ is the action taken, $R_t$ is the reward for taking action $A_t$, and $S_{t+1}$ is the next state. Then choose a regression algorithm to train.

\textbf{(2) Initialization:} Set N to 0 and initialize $\hat{Q}_N$ to be a function equal to zero everywhere on the data space $S \times A$ a Cartesian product of state space and action space.

Until stopping conditions are reached, do:

\textbf{(3) Iterations:}

- $N \leftarrow N+1$

- Build the next training set \{$S_t,A_t,R_t+\lambda \text{max}_A \hat{Q}_{N-1}(S_{t+1},A)$\} using $\hat{Q}_{N-1}$

- Continue using the regression algorithm to induce from the training set the function $\hat{Q}_N(S,A)$

Ernst et al., (2003) experimented with tree-based regression methods. Following (Ernst et al., 2013), we are going to use Extra-Trees algorithm in this paper due to the algorithm's desirable behaviors. Detailed explanation of extra-trees algorithm use and convergence in a fitted Q algorithm can be found in (Ernstet al., 2003).

\section{Trading with Heston model and fitted Q}
\subsection{The trading process}
In this study, we are going to adopt the trading process outlined in Ritter (2017). To make the trading systems realistic, we are limiting the agent's trading and holding size limits. The trade size $\Delta n_t$ in a single time interval is limited to at most $K$ round lots; and one round lot is equal to 100 shares. We also limit the holding position size to $M$ round lots. Therefore, the action space and the holding space are respectively:
\begin{equation}
\begin{aligned}
\mathcal{A} &= \text{LotSize} \ \cdot \ \{-K,-K+1,...,K\} \\
\mathcal{H} &= \{-M,-M+1,...,M\}
\end{aligned}
\end{equation}
with cardinalities $|\mathcal{A}|=2K+1$ and $|\mathcal{H}|=2M+1$. Another feature of the trading process is tick size, for example equals USD 0.01, so that all quoted prices are integer multiples of the tick size. After a trade is made, the agent is charged a spread cost of one tick size for any trade. Mathematically:
\begin{equation}
\text{SpreadCost}(\Delta n) = \text{TickSize} \ \cdot \ |\Delta n|
\end{equation}
We also assume that there is a permanent price impact with linear functional form: each round lot traded moves the price one tick and therefore leads to a cost of $|\Delta n_t| \times \text{TickSize/LotSize}$ per share traded. The impact dollar cost for all shares is:
\begin{equation}
\text{ImpactCost}(\Delta n) = (\Delta n)^2 \times \frac{\text{TickSize}}{\text{LotSize}}
\end{equation}
Using the cost functions above, we can get the profit/loss equation:
\begin{equation}
\text{PNL} = n_t(p_{t+1}-p_t) - \text{SpreadCost} - \text{ImpactCost}
\end{equation}
where $(p_{t+1}-p_t)$ is the change in price and $n_t$ is the number of stocks traded. We also follow Ritter (2017) in calculation of the reward function:
\begin{equation}
R_{t+1} = \text{PNL}_{t+1} - 0.5 \kappa (\text{PNL}_{t+1})^2
\end{equation}
This reward function comes from a mean-variance equivalence that is discussed in Ritter (2017). 
\subsection{Parameter estimation results}
As outlined in section 1, our first and second step in the multi-step training procedure are fitting real-world data using the model of our choice, the Heston model, and generating a large sample using the parameters obtained. We use interest rate $r=0.004$ percent, the approximate average daily one-year treasury bond interest rate of 2017, and prices of GSPC in 251 trading days of 2017 as our fitting data. Using the initial guess of $\sigma=0.1,\theta=0.1,\kappa=0.1,\rho=0.1$, we obtain a convergence graph of parameters:

\begin{center}
\includegraphics[width=1\textwidth]{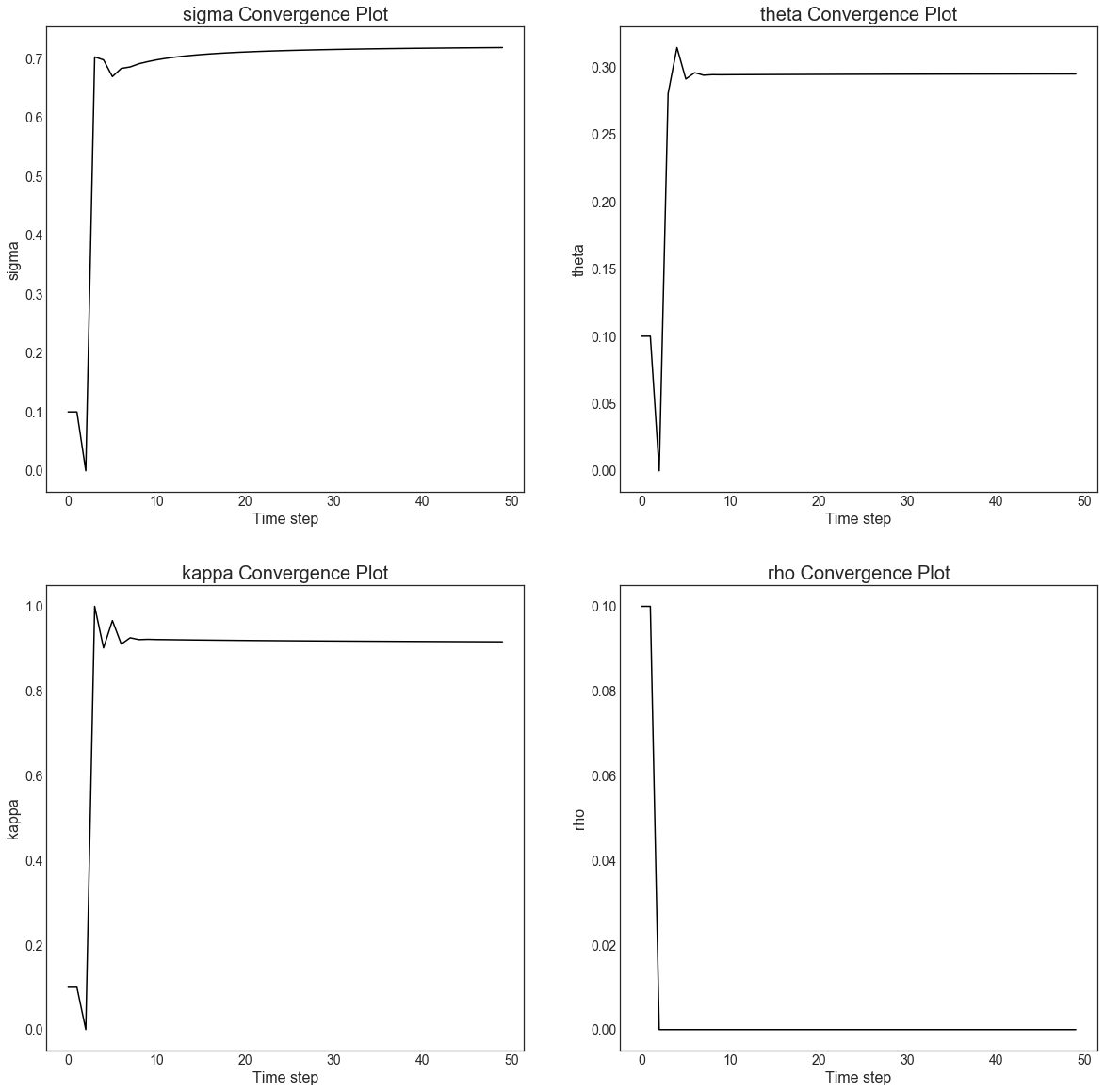}

Fig 1. Parameter convergence plot
\end{center}

With the parameters obtained, we can use them to generate a sample plot. Fig 2 below is a visualization of 100  generated sample paths using the fitted model and the stock price of GSPC in 2017. The accuracy of the extended Kalman filter and the pseudo-maximum likelihood estimation are discussed in (Javahari et al., 2013) and (Wang et al., 2017).

\begin{center}
\includegraphics[width=1\textwidth]{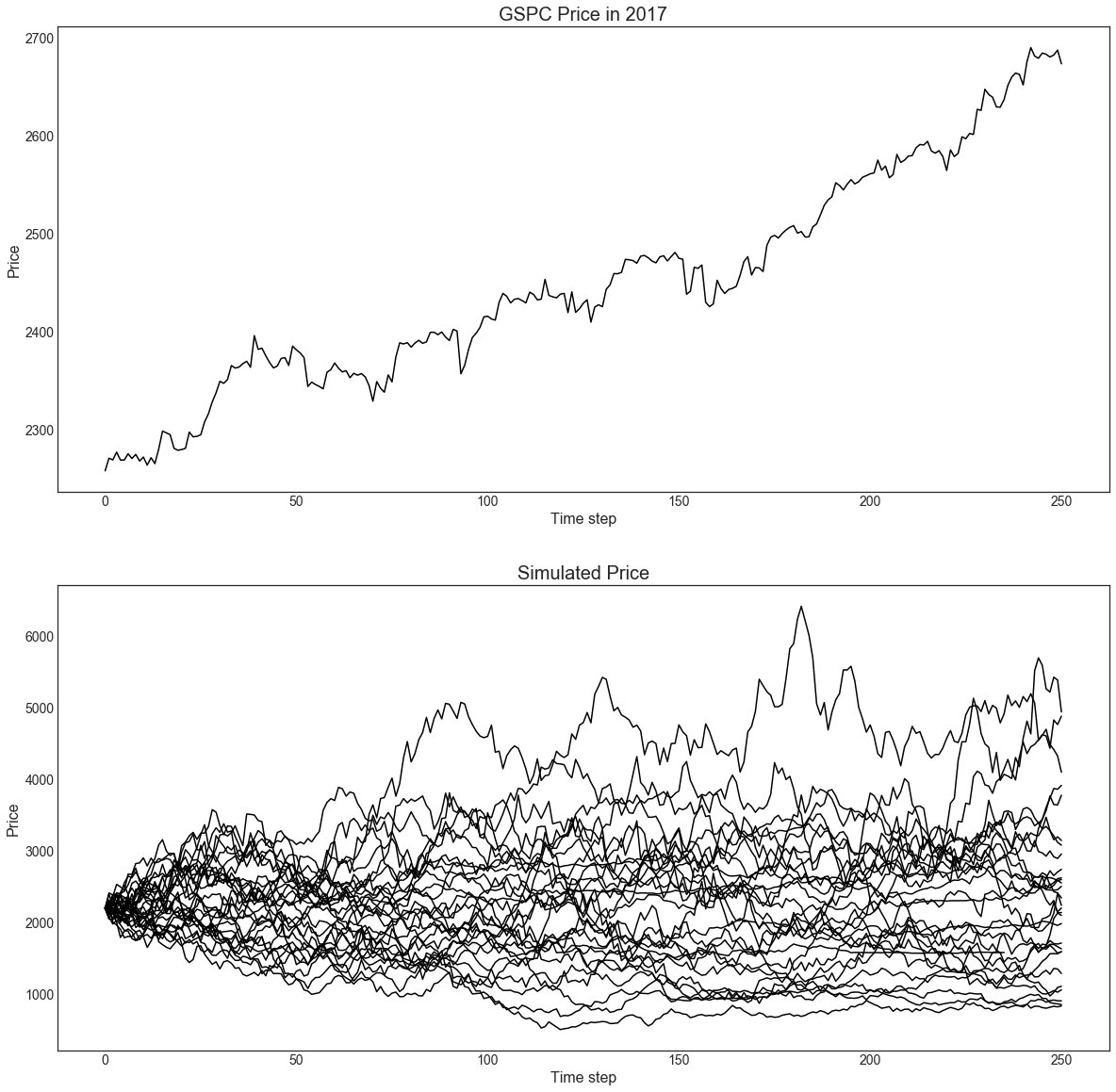}

Fig 2. Price plot and simulated path plot
\end{center}

For all our experiments, we are using a path of $10^5$ to train our agents. We define our trading and holding limit with $K=5, M=10$ respectively. Then the state space is the Cartesian product $\mathcal{S}=\mathcal{H} \times \mathcal{P}$. We pick $\kappa = 10^{-4}$ for the parameter of the reward function, and $\lambda=0.999$ for our $Q$-function discounted rate. The regressor we choose is the extra-tree regressor, an ensemble of 10 trees minimizing mean-square error. For parameters of the ensemble, we pick the minimum of 5 sample splits and 5 sample leafs. More about the importance of parameters of the extra-tree regressor can be found in (Earnst et al., 2003). Though the parameter choices can be important in machine learning, in our experiments, we are not doing so to expedite training.

As mentioned in (Ernst et al., 2003), using extra-tree regressor doesn't guarantee convergence, but it doesn't diverge to infinity and has been shown to deliver better results than using other converging regressors. After $10^4$ fitted Q iterations, we stop training to use the agent for out-of-sample trading test. 

\subsection{Trading experiments and results}

Our first experiment is similar to that of Ritter (2017). To test that an agent can be trained to trade profitably, we train agents using a stock return dynamic that allows for arbitrage. The paths generated for training and testing will include a mean-reverting mechanism to the stock price process. The possibility of arbitrage has been shown in Ritter (2017). However, different from Ritter (2017), our experiment doesn't use the Q-matrix and instead uses the fitted-Q algorithm whose parameters and estimators are discussed in prior sections. Furthermore, to illustrate that the fitted-Q iteration algorithm is capable of working with continuous values and higher dimension, we also include the volatility of stock price as one of the inputs besides stock price and stock position. We train 100 different agents by using 100 different generated paths of $10^5$ steps and then test the agents using another set of 100 different paths of $10^5$ steps.

\begin{center}
\includegraphics[width=1\textwidth]{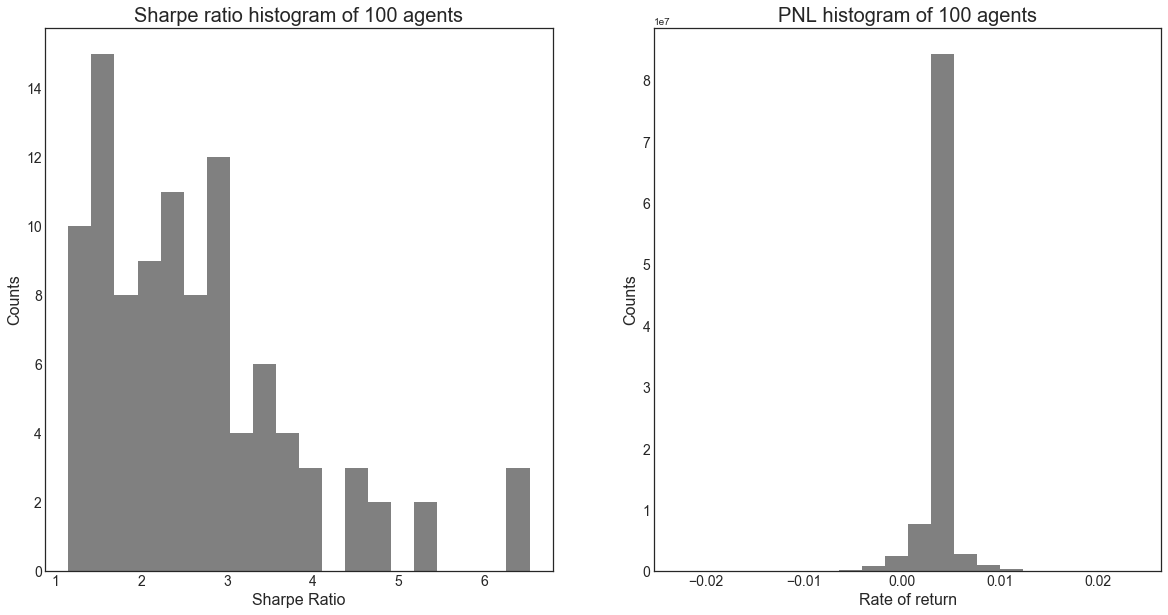}

Fig 3. Agents' Sharpe ratio and PNL plots in arbitrage environment
\end{center}

Fig 3 includes the histograms of the Sharpe ratio and rate of return obtained by the 100 agents. The average rate of return is $0.301\%$, with standard deviation of $0.15\%$. The average Sharpe ratio of the 100 agents is $2.64$ with standard deviation of $1.18$. As noted, profitability is expected in an environment that allows for arbitrage opportunity. However, high and stochastic volatility of the system have hindered the agents' trading and therefore the agents performs slightly worse than they perform in a simulated environment where volatility is constant. To help the agents learn more about the volatility process, more iterations and training sample can be included.

Our second experiment tests our agents on real life data. 450 agents are trained using the stock prices of 450 different S\&P 500 stocks. The experiment process for each of the 450 agents is as follow: (1) Fit the parameters of a Heston model using stock prices from the 2-year period of 2010-2012. (2) Create a sample path of $10^5$ steps. (3) Train the agent using the sampled path. (4) Record trade results of the agent using stock prices of the 5-year period from 2012 to 2017.

\begin{center}
\includegraphics[width=1\textwidth]{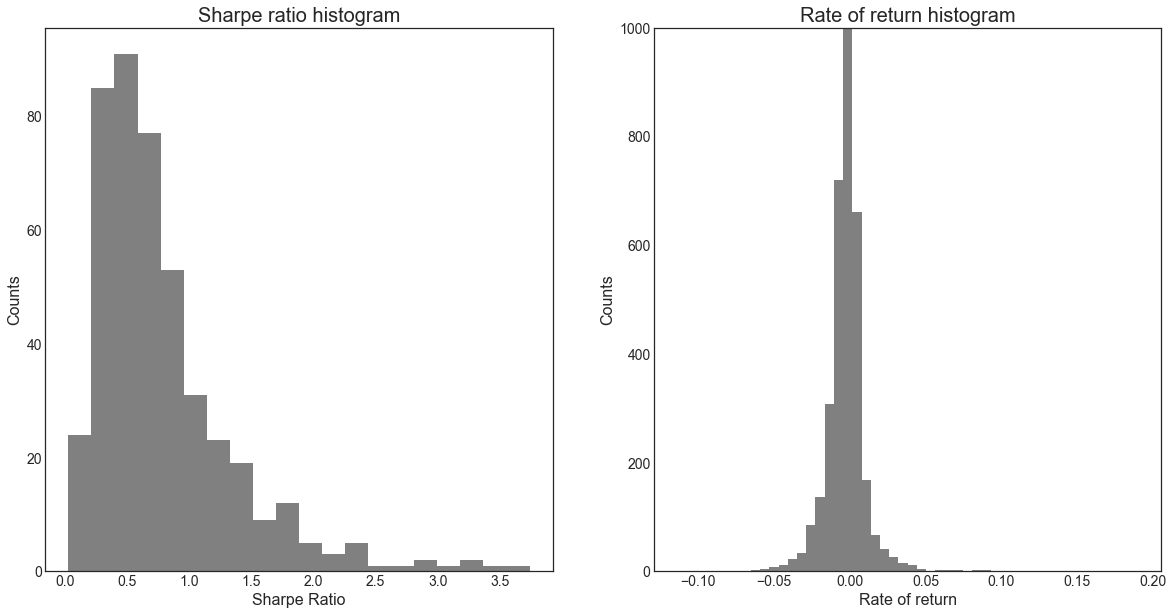}

Fig 4. Agents' Sharpe ratio and PNL plots with real life data
\end{center}

Fig 4 shows the trading results for the agents on real-life stock prices. The mean of 450 agents' Sharpe ratios is $0.785$ with standard deviation of $0.568$. We observe that the agents are not performing well in a risk-adjusted sense. The most common rate of return is $0.0\%$ as the agent learns about trading cost and therefore is hesitant to trade. The mean rate of return is $0.0282\%$ with standard deviation of $0.106\%$. To improve the agents, we can include other relevant variables with predictive value as well as increase the number of sample steps and number of iterations.

\section{Conclusions}
In this paper, we have explored the use of fitted Q iteration algorithm and Heston model in trading. A proof of concept, this paper hasn't fully shown the full possibilities of using fitted Q iteration in trading. More experiments can be conducted with increased environment and action size to include more economic or financial data to utilize the benefits of the regressors in fitted Q iteration. Variables such as volatility of the market, price of international indexes, interest rate forecast, and sentimental variables can be included and experimented to see whether better agents can be built.

In addition, one might test with different regressors other than the extra tree regressors to determine the difference among the results as well as the most optimal ones to be used for trading. For instance, using multi layer perception regression or extending the fitted Q iteration to a recurrent neural network can be interesting experiments. Moreover, one can also test the benefits of parameters tuning between every fitted Q iteration in order to determine the importance of parameters tuning in fitted Q iteration for trading.

Also, fitted Q iteration can also be tested for several other financial problems such as the portfolio management problem. Furthermore, according to neoclassical finance theory, no investor should hold a portfolio that does not maximize expected utility of final wealth. However, this fact is not reflected in this paper. Future paper can include different reward functions in order to reflect neoclassical finance theory better.

\end{document}